\documentclass[12pt,a4paper]{article}
\usepackage{amsfonts}
\usepackage{graphicx}
\usepackage{amssymb}
\usepackage{textcomp}
\everymath{\displaystyle}
\begin{document}
\title{Topological stability of broken symmetry on fuzzy spheres}
\author{S. Digal \thanks{digal@imsc.res.in} and T. R. Govindarajan \thanks{trg@imsc.res.in} \\
The Institute of Mathematical Sciences, C. I. T. Campus,\\
Taramani, Chennai-600113, India}
\maketitle

\begin{abstract}
We study the spontaneous symmetry breaking of $O(3)$ scalar field on a 
fuzzy sphere $S_F^2$. We find that the fluctuations in the background of topological
configurations are finite. This is in 
contrast to the fluctuations around a uniform configuration which diverge, 
due to Mermin-Wagner-Hohenberg-Coleman theorem, 
leading to the decay of the condensate.
Interesting implications of enhanced topological stability of the configurations are 
pointed out. 
\end{abstract}

\section{Introduction}

 Field theories on non-commutative geometries are inherently 
 non-local\cite{hoppe,madore,balbook,pinzul,balpinzul}. 
This feature gives rise to novel behaviour  such as the mixing of
infrared and ultraviolet scales. These lead to 
 new ground states with spatially varying condensates\cite{gubser,ambjorn,bietenholz1,bietenholz2}. 
Many  non-perturbative studies have established that fuzzy spaces such as 
 Groenewold-Moyal plane, fuzzy spheres allow for the formation of stable non-uniform 
 condensates as ground states\cite{xavier,denjoe,flores,denjoeydri,medinathesis,panero1,panero2,digal1,digal2,medina}. 
Nonlocality plays an essential role in 
 all these. Exploring implications of this nonlocal nature of field 
 theories is one of the intensive research activity in the field of 
 noncommutative physics today. 

In the past, we have analysed  the spontaneous breakdown of global symmetries 
and the corresponding behaviour of Goldstone modes through simulations\cite{digal1,digal2}. 
The main aim of this paper is to study the interplay of SSB and topological features of the 
sigma models on  fuzzy spheres.  This issue 
 is important because the Mermin-Wagner-Hohenberg-Coleman (MWHC) theorem states that there 
 can be no SSB of continuous symmetry on 2-dimensional commutative 
 spaces\cite{mermin,hohenberg,coleman,paolo}. This theorem was established rigorously by
summing up all the infrared contributions in the context of field theories
with local interactions. The resulting diverging fluctuations of the
Goldstone modes prevent the symmetry from being spontaneously broken.
 No SSB also implies no  magnetisation. There is no obvious generalisation of 
 the MWHC theorem for the non-commutative spaces, since the theorem relies 
strongly on the locality of interactions. 
 Gubser and Sondhi, \cite{gubser}, have shown that the Goldstone mode 
 fluctuations become severe in the non-commutative case. These authors 
 infer, after analysing the fluctuations in the background of a uniform condensate, 
 that a condensate with  only zero 
 momentum/angular momentum mode are unstable. However
 non-commutative spaces admit non-uniform solutions  
 and one can ask the question what happens to the stability of
 these configurations. Non-uniform condensates naturally have an infra-red 
cut-off for the fluctuations. This cut-off can soften the otherwise 
divergent contributions of the  Goldstone modes. It would be desirable to 
compute the fluctuations by analytical methods as was done for the MWHC 
theorem. Only few models allow for exact treatment of SSB even in the
commutative spacetime. It is difficult to analytically sum the fluctuations 
in the present case. But in the absence of such a study numerical simulations
can provide answers in the present context.

In a previous work of ours, we explored these issues for $U(1)$ symmetry using numerical 
 simulations \cite{digal2}. As expected,  we found that the uniform 
 condensates are unstable. But the condensates with higher angular momentum 
components still survived the fluctuations. However when the thermodynamic
 and continuum limits were taken, the flutuations in the momentum mode above the cut-off 
 increased. The Goldstone modes 
 were capable of destroying the low lying modes of the non-uniform condensate.

 In this paper we analyse the $O(3)$ symmetry on fuzzy sphere. 
Nonlinear sigma models on fuzzy spheres have been studied in the literature
exploring topological features \cite{trghari,balimmirzi}. We
 expect that if we repeat the calculations of $U(1)$ for this case 
 the results will be more or less similar. But the $O(3)$ case is
 significantly different from our previous study in an important way. In this 
 case there are non-uniform configurations which are topologically
 stable. For example the hedgehog type configuration. A condensate which is
 topologically stable can be unstable against fluctuations with energies 
 of the order of the system size. So we expect that Goldstone
 modes whose energy goes as logarithm of the system size may not be
 able to destroy the hedgehog condensates. To settle this, we 
 calculate the fluctuations around such configurations via numerical
 simulations. Our numerical results show, that indeed, topological 
 stability and nonlocal interactions make the lowest non-zero mode of 
 the condensate stable.
  
 This paper is organised as follows: In Sec.2 we describe the model 
 and the non-trivial topological configurations. The numerical simulations
 and the results are discussed in Sec.3. In Sec.4 we provide our
 conclusions.

\section{ $O(3)$ Model and Topological Condensates on $S_F^2$}

The fuzzy sphere is described by the coordinates $X_i$ satisfying the 
following $SU(2)$  algebra,
\begin{equation}
\left[X_i,X_j\right]=\frac{i\alpha \epsilon_{ij}{^{k}}X_k}{{\sqrt{j(j+1)}}},~~~
\sum_i X_i^2=R^2.
\end{equation}

\noindent Here $R$ is the radius of the fuzzy sphere. One can choose the
angular momentum $j~=~\frac{N}{2}$ representation for the fuzzy sphere. 

With this choice the most general action for $O(3)$ hermitian 
scalar field $\Phi_i$ $(i=1,2,3)$ upto quartic interactions takes the form,
\begin{equation}
S(\Phi) = \frac{4\pi}{N}\,{\rm Tr}\left[
\sum_i |\left[L_i,\Phi\right]|^2
+R^2\left(r|\Phi|^2+ i\beta \epsilon_{ijk}\Phi_i\Phi_j\Phi_k+\lambda(|\Phi|^2)^2+ 
\mu|\left[\Phi_i,\Phi_j\right]|^2\right)\right] 
\label{action}
\end{equation}

In the mean-field the above theory admits a uniform condensate for $r<0$.
However as discussed above the fluctuations of the Goldstone mode render
this solution unstable. Apart from the uniform condensate the above model
admits many meta-stable solutions. 
To simplify our arguments we first consider the case $\beta~=~0, \mu~=~0$:

The number of meta stable configurations
increases with the matrix size of $\Phi$. Among them  we are interested on those
which are stable due to topological obstructions. For example,
\begin{equation}
\Phi_i = ~\alpha L_i,~~ \rm{with} ~~\alpha = \sqrt{\frac{\frac{2|r|}{\lambda}}{N^2-1}}
\label{winding}
\end{equation} 

The analog of this configuration in continuum space is the hedgehog 
configuration where the $O(3)$ spin vector on the sphere is pointing 
radially outward. The spin is parallel to the position vector on
the sphere. This configuration is topologically stable as  
it cannot be smoothly deformed to a uniform one. 
Similarly the above configuration cannot be smoothly deformed to  
$\Phi~=~I$ which is also a solution. The above configuration corresponds to a winding one 
map from the physical space $S_F^2$ to the vacuum manifold which is 
$S_F^2$. All topologically stable configurations in the continuum limit,
can be characterised by the second homotopy group $\Pi_2(S^2)$. 
For a discussion on topological classification of the maps $S_F^2\longrightarrow S_F^2$
see \cite{trghari,balimmirzi}.

When we include the other terms in the action we have more parametric choices 
for the topological configurations. For example when $\beta \neq 0,~ \rm{but}~ \mu = 0$
we have 

\begin{equation}
\Phi_i = \alpha L_i, \rm{with} ~~\alpha = \frac{3\beta~\pm~\sqrt{9\beta^2~-32r\lambda~L^2}}
{8\lambda~L^2}
\label{newwinding}
\end{equation}

Then Eq.(\ref{newwinding}) shows clearly the possible existance of  topological 
vacuum configurations even when $r \geq 0$. For example, when $r~=~0$
we have 
\begin{equation}
\Phi_i = \alpha L_i, \rm{with} ~~\alpha ~= ~\frac{3\beta}{\lambda~(N^2~-~1)}
\label{newwinding1}
\end{equation}
While it is important to study the topological stability in such a general situation
the results  are not characteristically different from $\beta = 0$. Here we focus 
on $\beta = 0$ in this paper for simulations but discuss general cases. 

To study the net effect of topological nature of the background 
configuration and non-locality on fluctuations, we consider only the 
winding number one configuration which is given in Eq.(\ref{winding}).
As mentioned our plan is to compute these
fluctuations numerically. Even before computing the fluctuations
one can make some general remarks about the behaviour of the
fluctuations \cite{digal2}. The effect of nonlocality basically 
provides a non-zero mass $O(\frac{\alpha}{N})$ to the Goldstone mode fluctuations. This puts
an infrared cut-off for the fluctutations. From our previous study 
\cite{digal2} it seems that this mass/cutoff is mode dependent, as only 
higher modes of the condensate survived the fluctuations. As we will see
from our results the combined effect of topology and nonlocality, the infrared
cut-off  drastically  reduce the contribution of the fluctuations. In the next 
section we describe our numerical simulations. 

\section{Numerical simulations}

We use ``pseudo-heat bath" updating method in our numerical simulations
which is described in detail in our earliar papers \cite{digal1,digal2}. 
In our method, for each choice of parameters, we choose an initial 
configuration given by Eq. (\ref{winding}). Fluctuations around this 
configuration are then generated by the above updating method. 
Since this configuration is a variational solution to minimising the
classical action, it will thermalise as we update/include the thermal
fluctuations. Once the initial configuration is thermalised we compute 
the observable M. We make measurements
after every 10 updates of the entire matrix. We also use over-relaxation to 
reduce the auto-correlation of the configurations generated in the 
Monte-Carlo history.

In a numerical simulation, the condensate  will not maintain
its exact form as in Eq. (\ref{winding}) along the Monte-Carlo history. The configuration
can evolve into different random $SU(2)$ rotated configurations of Eq.(\ref{winding}) 
as we keep updating it. To overcome this, one needs to rotate the configuration at
each step of the Monte-Carlo history so that the configuration takes the 
form of Eq. (\ref{winding}). But this is a difficult and time consuming task. 
On the other hand
one can have an observable made of $\Phi_i'$s which is invariant under the 
$SU(2)$ rotations, e.g basis independent. For this purpose we define the 
following observable,
\begin{equation}
A_{ij} = \frac{1}{N^2} Tr(L_i\Phi_j), M = \sqrt{A^\dag A}
\end{equation}
$M$ projects out the $l=1$ angular momentum mode. Note that the initial
configuration in Eq. (\ref{winding}) projects out only the $l=1$ mode. Analysing the
statistical behavior of $M$ will give us a definite conclusion about the stability of the
initial configuration. We mention here that
$Tr(\sum_i \Phi_i^2)$ is also an $SU(2)$ invariant. But the information on the 
amplitudes of different $l$ modes gets lost in this form. Also comparatively
the observable $M$ may serve as an order parameter in the case of any 
phase transition of the hedgehog configuration to $\Phi_i = 0$ at high 
temperatures. We mention here that $l=1$ is the lowest possible stable mode, 
as $l=0$ mode will be unstable. One can consider configurations with higher 
winding, instead of Eq.(\ref{winding}), however we expect them to be more stable than 
the $l=1$ condensate. This is because the infrared cut off will rise
with higher winding configurations.

For practical reasons, the size of the matrix $N$, in other words size of
the resolution scale is finite. So there are usually finite volume $(R,N)$
effects. So a non-vanishing condensate $\Phi_i$ does not mean 
there is SSB. One needs to define suitable observable dependent on $\Phi_i$ 
which should scale with $(R,N)$ appropriately in the thermodynamic limit
$(N \rightarrow \infty,R \rightarrow \infty)$ to conclude anything. Now there are 
two possible thermodynamic limits. If in the thermodynamic limit the ratio 
$\frac{R^2}{N}$ does not vanish then the space is described by a non-commutative algebra. 
This limit  is of interest to us, as  we expect that the MWHC theorem will hold good
in the commutative thermodynamic limit.

\begin{figure}[hbt!]
\begin{center}
{\label{fig1a}\includegraphics[scale=0.59]{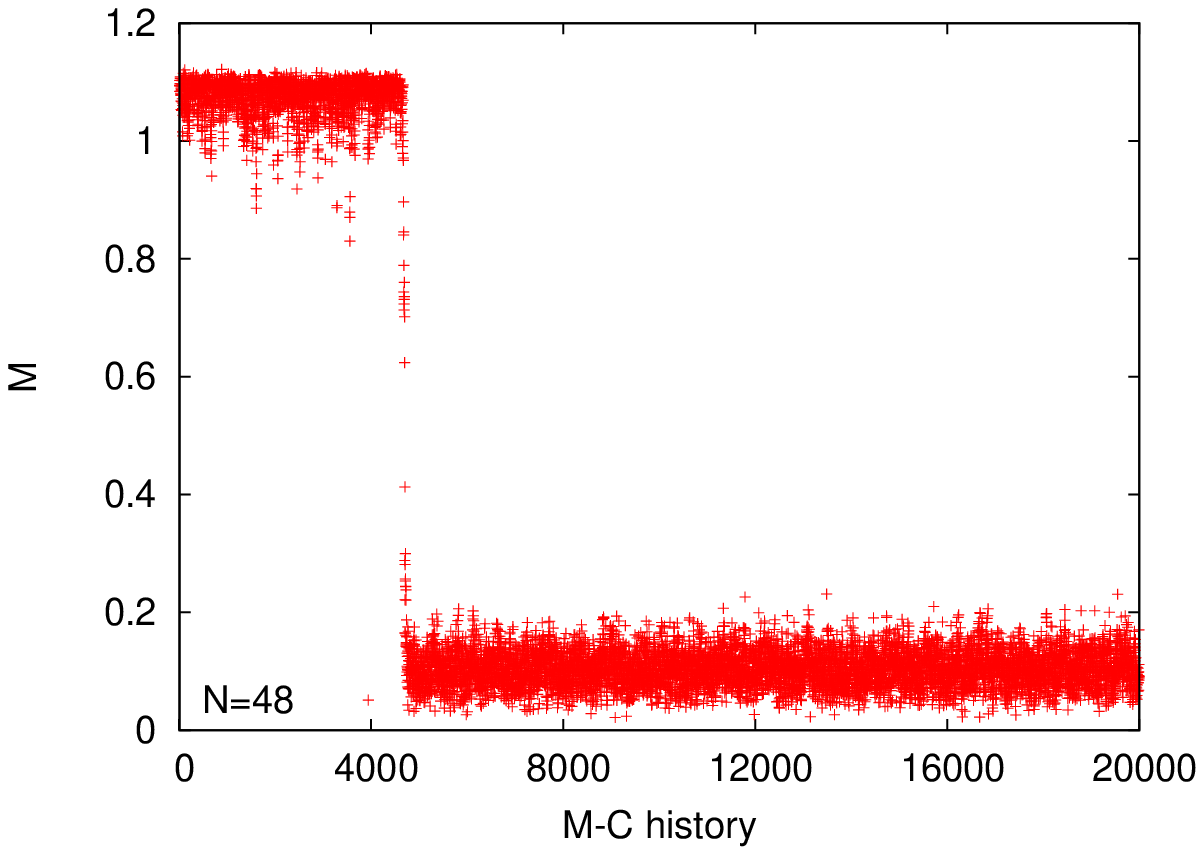}}
\caption{$M$ vs Monte Carlo history for $N=48$}
{\label{fig1b}\includegraphics[scale=0.59]{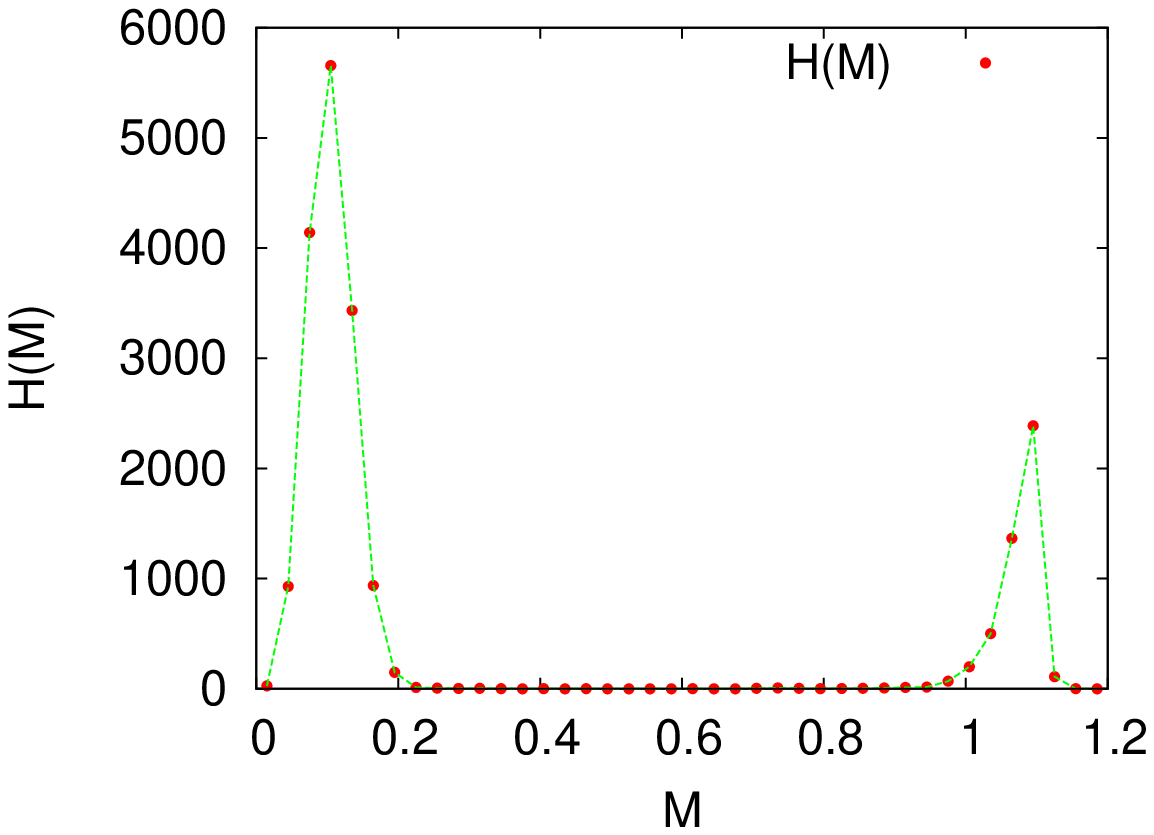}}
\caption{Histogram H(M) for $N=48$}
\end{center}
\end{figure}

\section{Results and discussions}

In our calculations we fix $\frac{R^2}{N} = 10$, $r = -8$. For simplicity
we take $\lambda_1 = 0.25$. With this choice of parameters
we do our simulations for five different sizes of the $\Phi$
matrices, $N=48, 56, 64, 78, 96$. FIG.1 gives a typical Monte-Carlo
history of our simulations for $N=48$.

In FIG.1 $M$ fluctuates around a value close to the initial value. 
Then $M$ suddenly jumps to a small value and settles down. A histogram $H(M)$ of
$M$ clearly shows two peaks $M$ as seen in FIG.2. The peak on the
left has large $l=0$ and small $l=1$ component. The peak at higher 
value of $M$ has large $l=1$ component and small $l=0$. This peak is 
close to the value of the initial configuration. So in this state 
fluctuations modify the initial configuration slightly and retain its 
topological nature. In our Monte-Carlo history we observed the $l=1$ state
decaying to $l=0$ state but not vice-versa. This implies that due to 
finite volume effects, the uniform condensate is more stable than our 
initial hedgehog configuration for this case of $N=48$. 

To study the stability of the $l=1$ configuration we considered both the 
commutative and non-commutative limit. For the commutative limit we fixed 
$R^2$ and considered higher values of $N$. We did not observe any
change in the distribution of $M$ in the $l=1$ state. The average value,
and the fluctuations of $M$ remain almost the same as we go from 
$N=48 \rightarrow 64$, as can be seen in FIG.3. As for $N=48$ the $l=1$
configuration also decays for $N=64$. This result suggests that the 
$l=1$ topological configuration is not stable in the commutative
continuum limit as expected. 

\begin{figure}[hbt!]
\begin{center}
{\label{fig2a}\includegraphics[scale=0.59]{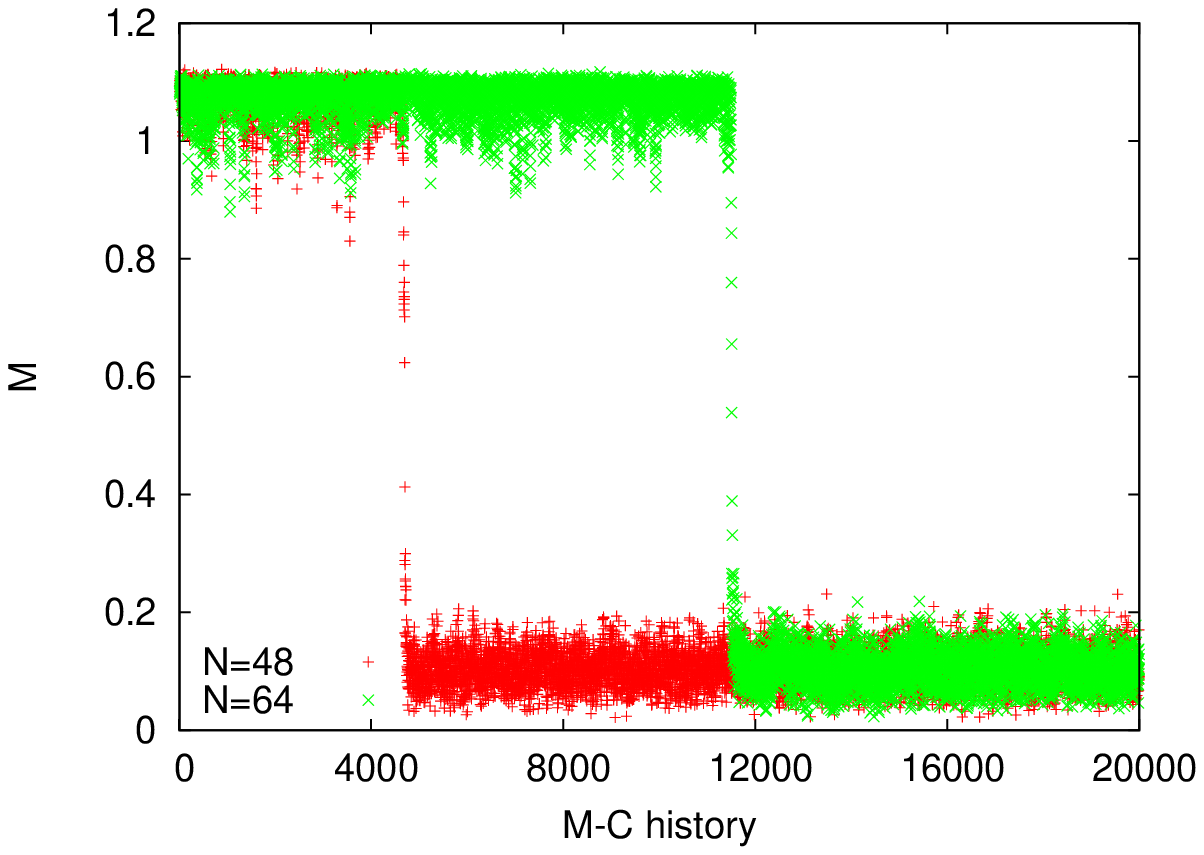}}
\caption{M-C history for fixed $R^2$}
{\label{fig2b}\includegraphics[scale=0.59]{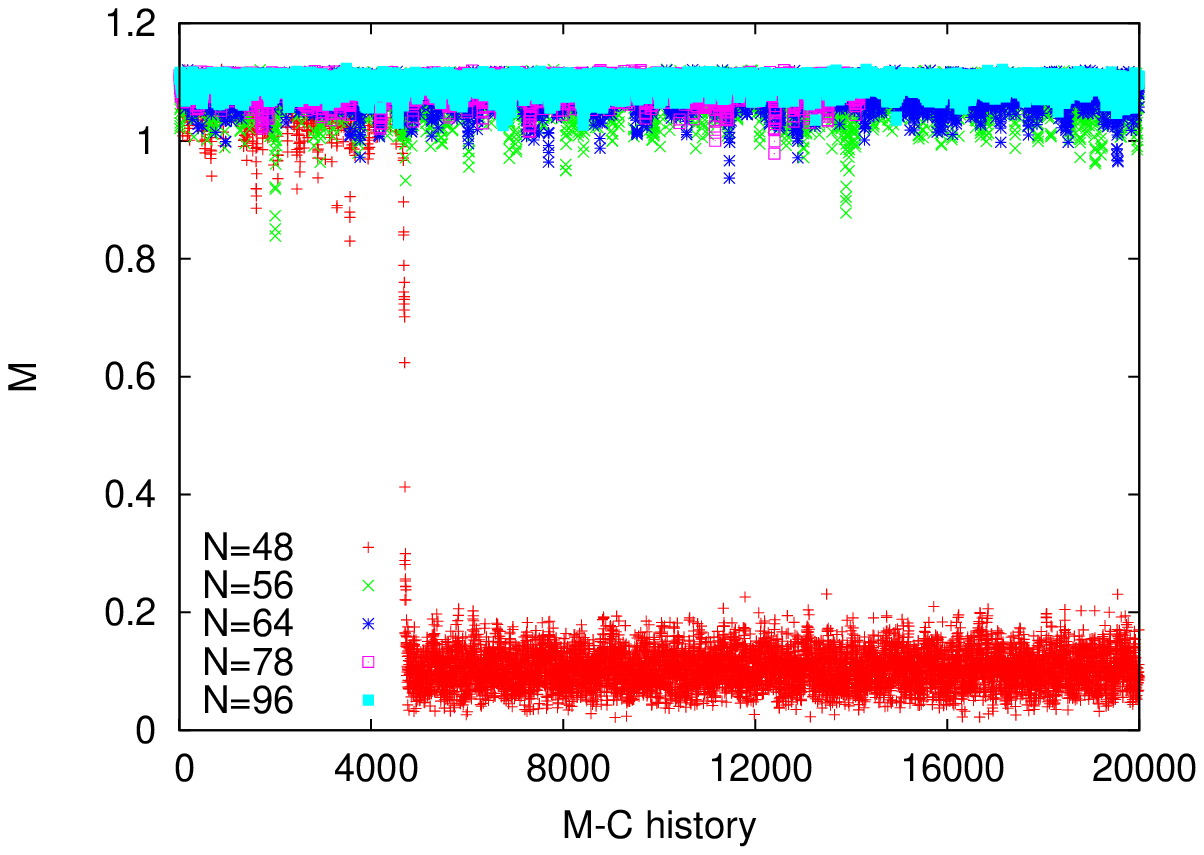}}
\caption{M-C history for fixed $R^2/N$}
\end{center}
\end{figure}

There is a complete change in the behavior as we consider the
non-commutative limit, i.e fixed $\frac{R^2}{N}$ as we
increase $N$. Except for the lowest $N=48$ the $l=1$ state did not decay 
during the entire run for higher $N$. In FIG.4 we show the Monte-Carlo 
history of $N=48,64,96$. Unlike the commutative limit, the fluctuations of 
$M$ decrease with $N$. This can be clearly seen in FIG.4. In FIG.5, we
give the  average value of $M$ as a function of $N$. 
The average value of $M$ increases slightly
with $N$, with the variation decreasing with $N$. This suggests $M$ will
reach a finite value in the continuum limit. We also compute the fluctuations
of $M$ to see any possible scaling with the cut-off $N$.
In FIG.6, we show $\chi~=~\left< M^2 \right > - \left< M \right> ^2$
in the $l=1$ state. The solid curve represents a fit, $f(N) \sim N^\alpha$
with $\alpha\sim -4.$. This clearly suggest that the $l=1$ state is stable 
in the $N \rightarrow \infty$ leading to spontaneous 
breaking of the $O(3)$ symmetry. 

\begin{figure}[hbt!]
\begin{center}
{\label{fig3a}\includegraphics[scale=0.59]{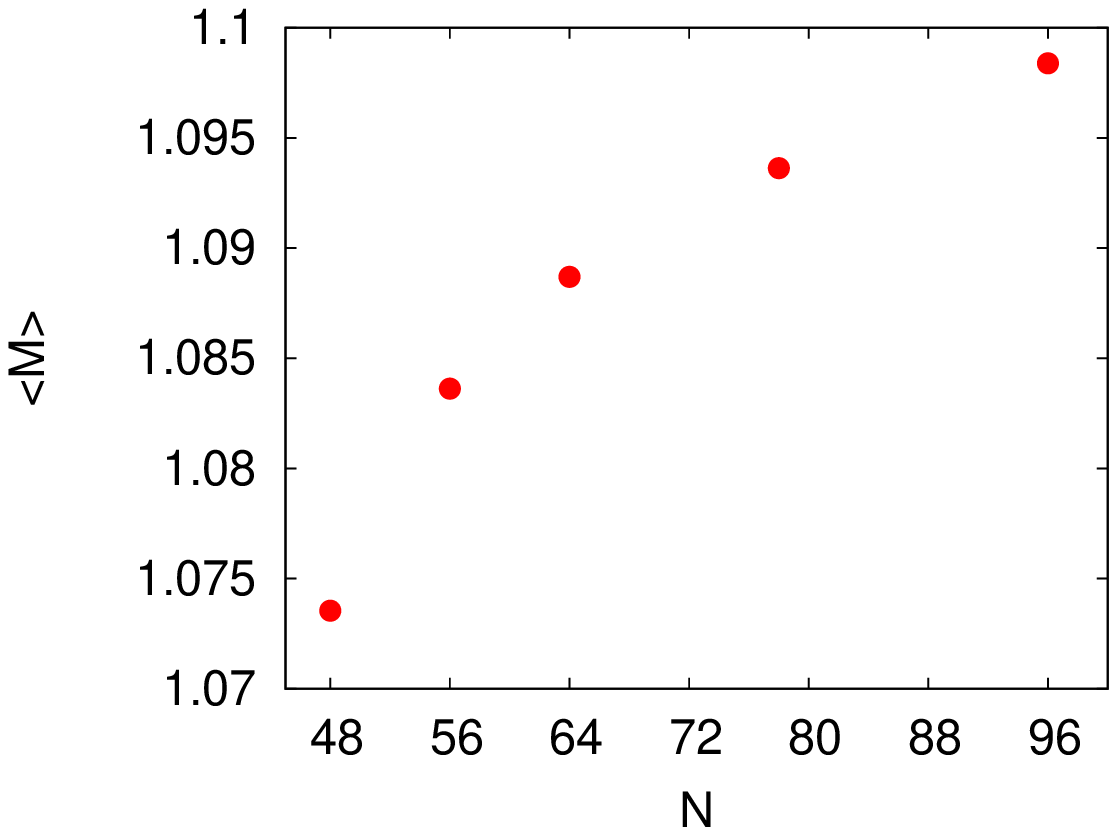}}
\caption{$\left<M\right>$  vs  $N$}
{\label{fig3b}\includegraphics[scale=0.59]{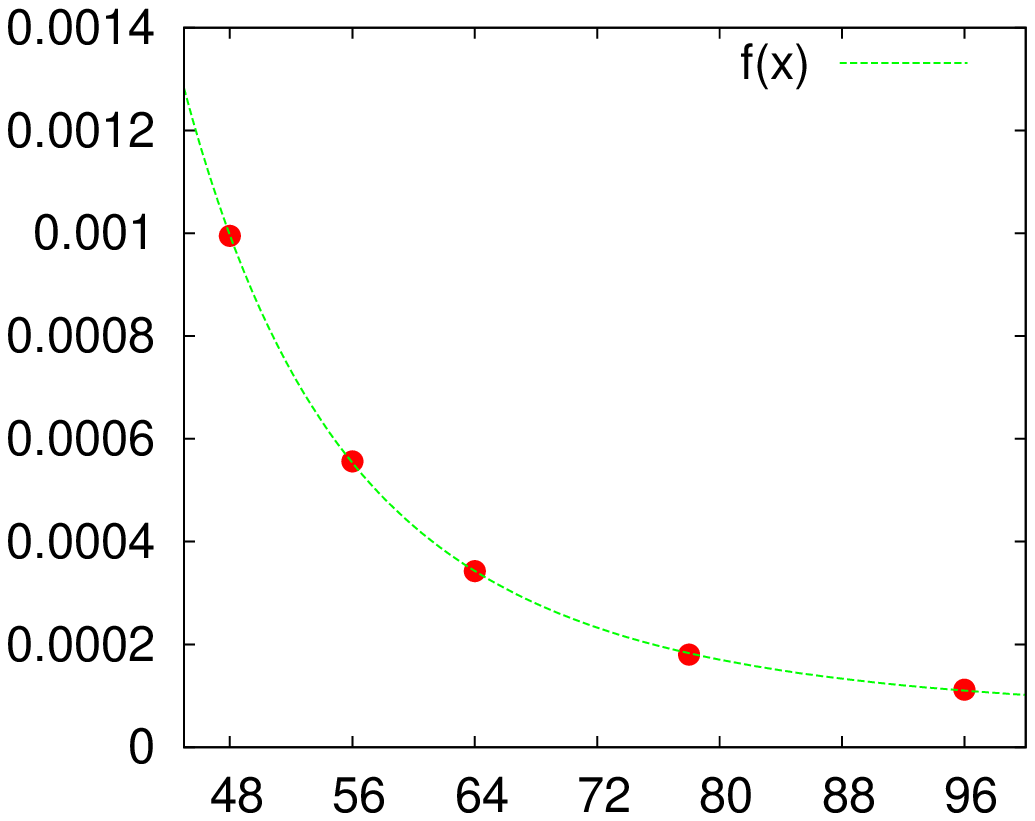}}
\caption{$\chi$ vs $N$}
\end{center}
\end{figure}

We also mention here that one can start with an initial uniform $l=0$ 
configuration and consider fluctuations. We expect that the results be similar 
to that in ref.\cite{digal2}. In ref.\cite{digal2} it was
found that only the highest mode $l = \frac{N-1}{2}$ condenses. The fact 
that we find the $l=1$ mode stable clearly shows that the topological nature 
of the initial configuration complements the effect of non-locality. These two
effects drastically reduce the fluctuations. 

\section{Conclusions}

In this analysis we have shown that topologically non-trivial 
configurations on the fuzzy sphere
avoid the MWHC theorem much more dramtically than the non-topological symmetry breaking. 
The mass gap or the infrared cut-off in this case is
large enough to render the fluctuations of the Goldstone modes finite. On the
other hand for non-topological condensates the Goldstone modes are large
enough to destroy almost all the modes except the few highest modes.
It is interesting to scale $\lambda$ and $r$ as we go to continuum limit.
The limit of noncommutative continuum geometry is always expected to 
maintain SSB and  stable solitons as there is obstruction to MWHC theorem from 
nonlocality. This analysis will be presented elsewhere. 

We have presented the simulations wherein the cubic Chern-Simons (CS) term is absent
in the action Eq (\ref{action}). 
The Chern-Simons term allows topological solitons even when the 
quadratic mass term is positive upto some value. 
Interestingly with CS term the configuration $\phi_i ~=~ \alpha~L_i$ 
is preferred over symmetric solution.
On the otherhand, it is not expected 
to alter the picture of topological stability of the solutions. This term plays an important 
role in the emergent geometry in NC fuzzy spaces \cite{steinacker,ydri}.
What we find here in the simulations is that even in the absence of CS term, emergent fuzzy spaces can be stable.
The stabilty of higher dimensional fuzzy spaces like $CP_F^2$ are of significance in this
context \cite{dolan}. The implications of this stability for extra-dimensional fuzzy spaces will 
be considered later.

\section{Acknowledgment}
This work was done as a part of the CEFIPRA/IFCPAR project. 
We acknowledge the support from CEFIPRA/IFCPAR.
This was completed at  University of Tours when the authors visited
under Indo French collaborative project. We thank Dr Xavier Martin 
for the support.


\begin{thebibliography}{99}
\bibitem{hoppe} J. Hoppe, Ph.D. Thesis, MIT (Cambridge MA, 1982).
\bibitem{madore} J. Madore, Class. and Quant. Grav. {\bf 9}, 69 (1992),
\bibitem{balbook} A.P. Balachandran, S. Kurkcuoglu and S. Vaidya, arXiv: hep-th/0511114.
\bibitem{pinzul} A.P. Balachandran, T.R. Govindarajan and B. Ydri,
Mod. Phys. Lett. {\bf A15}, 1279 (2000) [hep-th/9911087].
\bibitem{balpinzul} A.P. Balachandran, A. Pinzul and B.A. Qureshi, JHEP
{\bf 0512}, 002 (2005) [hep-th/0506037].
\bibitem{gubser} S.S. Gubser and S.L. Sondhi, Nucl. Phys. {\bf B605},
395 (2001) [hep-th/0006119].
\bibitem{ambjorn} J. Ambjorn and S. Catterall, Phys. Lett. {\bf B549},
253 (2002) [hep-lat/0209106].
\bibitem{bietenholz1} W. Bietenholz, F. Hofheinz and J. Nishimura, 
Acta. Phys. Polon. {\bf B34}, 4711 (2003) [hep-th/0309216].
\bibitem{bietenholz2} W. Bietenholz, F. Hofheinz and J. Nishimura,
Nucl. Phys. Proc. Suppl. {\bf 129}, 865 (2004) [hep-th/0309182].
\bibitem{xavier} X. Martin, JHEP {\bf 0404}, 077 (2004) [hep-th/0402230].
\bibitem{denjoe} J. Medina, W. Bietenholz, F. Hofheinz and
D. O'Connor, PoS {\bf LAT2005}, 263 (2005) [hep-lat/0509162]. 
\bibitem{flores} F. G. Flores, D. O'Connor and X. Martin, PoS {\bf LAT2005}, 262 (2006) [hep-lat/0601012].
\bibitem{denjoeydri} D. O'Connor and B. Ydri, JHEP {\bf 0611}, 016 (2006) [hep-lat/0606013].
\bibitem{medinathesis} J. Medina,  Phd. thesis, arXiv: 0801.1284 [hep-th]. 
\bibitem{panero1} M. Panero, SIGMA {\bf 2}, 081 (2006) [hep-th/0609205]. 
\bibitem{panero2}JHEP {\bf 0705}, 082 (2007) [hep-th/0608202].
\bibitem{digal1} C.R. Das, S. Digal and T.R. Govindarajan, Mod. Phys. Lett. {\bf A23},
1781 (2008).
\bibitem{digal2} C.R. Das, S. Digal and T.R. Govindarajan, 
Mod. Phys. Letts {\bf A24} (2009) 2693; arXiv:0801.4479 [hep-th].
\bibitem{medina} J. Medina, W. Bietenholz and D. O'Connor,
JHEP {\bf 0804}, (2008) 041;  arXiv:0712.3366 [hep-th].
\bibitem{mermin} N.D. Mermin and H. Wagner, Phys. Rev. Lett {\bf 17},
1133 (1966).
\bibitem{hohenberg} P.C. Hohenberg, Phys. Rev. {\bf 158}, 383 (1967).
\bibitem{coleman} S.R. Coleman, Commun. Math. Phys. {\bf 31}, 259 (1973).
\bibitem{paolo} P. Castorina and D. Zappala, 
Phys. Rev {\bf D77} (2008) 027703; arXiv: 0711.2659 [hep-th].
\bibitem{trghari} T R Govindarajan and E Harikumar, Phys. Letts., {\bf 
B 602} (2004) 238.
\bibitem{balimmirzi} A P Balachandran and G Immirzi, Int Jour. Mod. Phys. {\bf
A19} (2004) 5237.
\bibitem{steinacker} Harold Steinacker, Nucl. Phys. {\bf B810} (2009) 1
\bibitem{ydri} Rodrigo-Delgadillo Blando, Denjoe O'Connor, B Ydri,
Phys. Rev. Letts, {\bf 100} (2008) 201601.
\bibitem{dolan} Brian P Dolan, Idrish Huet, Sean Murray, Denjoe O'Connor,
JHEP{\bf 0707} (2007) 007. 
\end{thebibliography}
\end{document}